\documentclass{article}

\usepackage{arxiv}

\usepackage[utf8]{inputenc} 
\usepackage[T1]{fontenc}    
\usepackage{hyperref}       
\usepackage{url}            
\usepackage{booktabs}       
\usepackage{amsfonts}       
\usepackage{nicefrac}       
\usepackage{microtype}      
\usepackage{lipsum}
\usepackage{graphicx}

\RequirePackage{amsmath,amsfonts,amssymb}
\RequirePackage[superscript,biblabel,nomove]{cite}
\newcommand{\etal}{\textit{et al}.}
\newcommand{\ie}{\textit{i}.\textit{e}.}
\newcommand{\eg}{\textit{e}.\textit{g}.}
\newcommand{\etc}{\textit{etc}.}
\newcommand{\viz}{\textit{viz}.}
\newcommand{\R}{\ensuremath{\mathbb{R}} }
\newcommand{\G}{\ensuremath{\mathcal{G}} }
\newcommand{\D}{\ensuremath{\mathcal{D}} }

\newcommand{\synth}{Syn$\mu$S}
\newcommand{\real}{Real$\mu$S}

\title{Deep learning for synthetic microstructure generation in a materials-by-design framework for heterogeneous energetic materials}

\author{
 Sehyun Chun \\
  Department of Industrial and Systems Engineering\\
  University of Iowa\\
  Iowa City, IA 52242 \\
  \texttt{sehyun-chun@uiowa.edu} \\
   \And
 Sidhartha Roy \\
  Department of Mechanical Engineering\\
  University of Iowa\\
  Iowa City, IA 52242 \\
  \texttt{sidhartha-roy@uiowa.edu} \\
  \And
 Yen Thi Nguyen \\
  Department of Mechanical Engineering\\
  University of Iowa\\
  Iowa City, IA 52242 \\
  \texttt{yenthi-nguyen@uiowa.edu} \\
  \And
 Joseph B. Choi \\
  Department of Industrial and Systems Engineering\\
  University of Iowa\\
  Iowa City, IA 52242 \\
  \texttt{boogun-choi@uiowa.edu} \\
  \And
 H.S. Udaykumar \\
  Department of Mechanical Engineering\\
  University of Iowa\\
  Iowa City, IA 52242 \\
  \texttt{hs-kumar@uiowa.edu} \\
  \And
 Stephen S. Baek \\
  Department of Industrial and Systems Engineering\\
  University of Iowa\\
  Iowa City, IA 52242 \\
  \texttt{stephen-baek@uiowa.edu} \\
}

\begin{document}
\maketitle
\begin{abstract}
The sensitivity of heterogeneous energetic (HE) materials (propellants, explosives, and pyrotechnics) is critically dependent on their microstructure. Initiation of chemical reactions occurs at hot spots due to energy localization at sites of porosities and other defects. Emerging multi-scale predictive models of HE response to loads account for the physics at the meso-scale, \ie{} at the scale of statistically representative clusters of particles and other features in the microstructure. Meso-scale physics is infused in machine-learned closure models informed by resolved meso-scale simulations. Since microstructures are stochastic, ensembles of meso-scale simulations are required to quantify hot spot ignition and growth and to develop models for microstructure-dependent energy deposition rates. We propose utilizing generative adversarial networks (GAN) to spawn ensembles of synthetic heterogeneous energetic material microstructures. The method generates qualitatively and quantitatively realistic microstructures by learning from images of HE microstructures. We show that the proposed GAN method also permits the generation of new morphologies, where the porosity distribution can be controlled and spatially manipulated. Such control paves the way for the design of novel microstructures to engineer HE materials for targeted performance in a materials-by-design framework. 
\end{abstract}


\section{Introduction}
Propellants, explosives, and pyrotechnics (collectively termed, ``energetic materials'') are key performance components in a wide range of applications, including solid rocket motors and munitions. Typically, these materials are mixtures of energetic organic crystals and materials such as plasticizers,\cite{DENNIS2019} metals,\cite{VARUNKUMAR2018} and other inclusions.\cite{Elbasuney2019,Zhen2019} Such heterogeneous energetic (HE) materials have complex, stochastic microstructures. The sensitivity of HEs to loading, \ie{} their tendency to detonate, is intimately linked to their stochastic micro-morphology.\cite{Perry_2017,Price1986} In fact, several micro-scale mechanisms,\cite{Field1992} such as collapse of voids/pores,\cite{Menikoff1999} inter-crystal friction,\cite{peterson2007} plastic deformation,\cite{winter1975} and shear-banding \cite{austin2015} play important roles in determining the sensitivity of HEs. All of these mechanisms contribute toward energy localization at micro-scale heterogeneities leading to a distribution of ``hot spots.'' Reactions are initiated at hot spots \cite{Field1992} and propagate outward, culminating in full-blown detonation under suitable conditions. Naturally, designers of devices containing such energetic materials are interested in predicting and controlling the initiation sensitivity of the HE materials, from the standpoint of safety as well as to precisely control and tailor the performance of myriad systems that rely on HEs.

The frontier in computational energetic materials research is to develop predictive multi-scale models to guide the design process of novel materials with tailored performance via microstructural control.\cite{bruck2007,muravyev2019} Predictive frameworks of energetic material response to loads are a matter of concerted current development by several groups worldwide;\cite{barnes2019, gambino2019, sen2018, saurel2017, wood2018, jackson2018, miller2020} such capabilities are needed to establish structure--property--performance (S--P--P) linkages as necessary precursors to materials-by-design of heterogeneous materials.\cite{bruck2007, goodman2017, hwa2017} The work presented in this paper is directed towards establishing a computational approach to relate stochastic microstructures of energetic materials to their observed performance. To achieve this overarching goal, \textit{in silico} experiments on an ensemble of stochastic microstructures have been performed in the previous work, \cite{rai2015,lee2019} to extract quantitative data from resolved meso-scale simulations in the form of surrogate models.\cite{nassar2019modeling,rai2019void} These models are used to close the macro-scale governing equations in a multi-scale predictive framework. The work in this paper presents a deep learning approach to generate ensembles of synthetic microstructures that can be used for simulations; the methodology also allows for designing new microstructures, paving the way for materials-by-design of energetic materials.\cite{muravyev2019,mcclain2019} Simulations on an ensemble of microstructures also facilitate uncertainty quantification (UQ) due to microstructural variability\cite{lee2019} \ie{} aleatory uncertainties associated with the inherently stochastic microstructure. The propagation of uncertainty across scales influences the overall prediction uncertainty of HE sensitivity at the macro-scale. Therefore, ensemble simulations performed on a sufficiently large set of synthetic microstructures will be a key enabling tool for the reliability-based design \cite{gaul2015, zhang2000, cho2014} of next-generation HEs.

\subsection{Need for ensemble simulations on synthetic microstructures}
In recent years, multi-scale, multi-physics models that establish structure--property linkages have begun to be developed for the response of HEs to shock and impact loading.\cite{roy2019} A key task in the multi-scale modeling workflow is the performance of ensembles of high resolution, high-fidelity simulations of the meso-scale reactive mechanics.\cite{rai2015} The goal of such simulations is to capture the essential physical ingredients and to quantify energy localization at micro-scale morphological features such as pores and crystal-crystal interfaces. A predominant mechanism for energy localization in the microstructure is the creation of hot spots which are formed due to the collapse of pores.\cite{menikoff2004pore} Pore collapse is a well-studied problem, both theoretically \cite{springer2018, levesque2015effect, rai2017collapse, rai2017high} and experimentally.\cite{bourne1992shock} Reactive simulations of the dynamics of pore collapse in microstructures extracted from images taken using scanning electron microscopy (SEM), x-ray computed tomography (XCT), \etc{} have also been performed.\cite{rai2015} However, since the microstructure is stochastic, a large ensemble of meso-scale calculations, using a large enough number of statistically representative microstructures is needed to extract statistically meaningful information from such simulations. However, computational modelers typically do not have access to large sets of imaged data for a variety of types of energetic materials, or even for various micro-morphologies of a single type of energetic material. Image acquisition is expensive, and imaged data may be distribution-sensitive and specific to limited formulations and material types. 

From a computational mechanics standpoint, it is extremely useful to have access to an ensemble of microstructural geometries so that simulations can be performed and the statistics of micro-morphologies can be correlated with measures of sensitivity. In the absence of a large database of microstructural images, computational scientists rely on generated synthetic microstructures\cite{jackson2018, alveen2013micromechanical, grabowski2019modelling, park2017three, vsavija2019modelling, groeber2014dream, barua2012energy} (hereinafter abbreviated as ``\synth{}'') as proxies to the real microstructures (hereinafter abbreviated as ``\real{}''); a large array of stochastic \synth{} that closely mimic the real sample must be created and used in \textit{in silico} experiments. Generation of realistic \synth{}, with adequate statistical representation and ability to control global and local features in a versatile and flexible computational framework will prove to be a key component in the materials-by-design process for precise and controlled performance of energetic materials.

\subsection{Previous approaches}

Generating an ensemble of stochastic \synth{} that stand in for the \real{} is a challenging task. In general, \synth{} can be generated using several different approaches.\cite{groeber2014dream, mandal2018generation} Of these, approaches based on shape descriptors have been used in the past; objects are inserted into a computational domain using shape packing algorithms constrained by global shape descriptors, such as volume fractions and particle size distributions. While packing algorithms \cite{mandal2018generation} can be used to generate microstructures with specified morphometric characteristics (\eg{}, porosity, particle size distributions), these methods are limited to regular/analytical shapes such as spheres, ellipsoids, and polygons.\cite{kumar2008reconstruction} It is also difficult to pack regular shapes for theoretical maximum densities (TMDs) significantly higher than the close-packing limit; typical pressed energetic materials of the type simulated in the current work can have TMD values greater than 90\%.\cite{Welle_2014} Another popular approach, particularly to achieve high TMDs, is to start with space-filling polygons via tessellation, followed by superposition of voids or other phases in the mixture.\cite{kim2016computational} In both packing and tessellation approaches, however, the microstructure fails to mimic the morphology of real samples for a broad class of heterogeneous materials; \ie{} flexibility as well as realism are lacking. In the particular context of energetic materials, both shape-packing and tessellation-based approaches\cite{jackson2018, kim2016computational} have been used to perform meso-scale simulations of the shock response of stochastic microstructures. However, the \synth{} generated by these previous approaches are still too ``ideal'', in that the range of possible void/defect/interface shapes are not well represented in these two approaches. This shortcoming is significant for two reasons: 

\begin{enumerate}
    \item \real{} possess features that contain non-ideal distributions of shape features, including outlier features such as large cracks and elongated, tortuous void structures. It has been shown experimentally that such structures possessing large surface-to-volume ratios and other shape characteristics play a significant role in energy localization and sensitivity.\cite{molek2020impact}
    \item Computational studies \cite{levesque2015effect, handley2018understanding, rai2017collapse} have shown that local features in microstructures, such as inter-void distances, void shapes and orientations are in fact key aspects of sensitivity that distinguish different classes of the same material. Therefore, \synth{} must mimic not only global features but also local structural characteristics to be useful in building predictive models of material performance and materials-by-design frameworks.
\end{enumerate}

Machine learning approaches are promising alternatives to overcome the limitations of shape-descriptor based approaches. In recent years, convolutional neural networks\cite{Lecun_2015} (CNN) have been used to learn patterns and textures from \real{} and generate \synth{} for a wide range of materials.
Among a variety of architectures, Li \etal{} \cite{Li_2018} proposed a method for \synth{} generation based on a general-purpose texture synthesis method in computer vision that uses transfer learning.\cite{gatys_2015} For a given \real{} input, their method generates a \synth{} having the same ``style'' as the input \real{} by minimizing the style difference, where the style of a microstructure is defined by the Gram matrix of feature maps produced by the CNN. However, the transfer learning method requires a \real{} as a reference to generate \synth{}. Hence, from the materials-by-design standpoint, material morphology can be explored only in the ``neighborhood'' of the existing \real{} samples. This poses a critical limitation in generating a large ensemble of microstructures that can span the space of candidate material morphologies. In another deep learning approach, Cang \etal{} \cite{Cang_2018} and Guo \etal{} \cite{Guo_2018} employed an encoder-decoder architecture to generate \synth{}. The encoder-decoder architecture develops a codified representation of micro-morphology by learning to compress the image pixels (encoder) and reconstruct it back to the original one (decoder). The code values learned by encoder-decoder networks parameterize micro-morphology, allowing the generation and manipulation of \synth{} by ``turning knobs,'' where the code values act as the knob control parameters. However, these networks tend to generate blurry images.


On the other hand, another deep learning-based method, the patch-based generative adversarial networks (GAN), can generate much sharper and crisper images as we will demonstrate later. Unlike the original GAN method, \cite{GAN_2014} patch-based GAN approaches evaluates the quality of generated images at different local regions, enforcing the details to be clearer and more realistic. Yang \etal{} \cite{Yang_2018} and Mosser \etal{} \cite{Mosser_2017} demonstrated that GANs are not only capable of generating realistic \synth{} but can also be used to continuously parameterize the micromorphology; this paves the way for smoothly varying the morphology to produce new microstructures. Fokina \etal{} \cite{fokina_2019} also confirmed the good performance of GANs in generating \synth{} of ALPORAS aluminum foam. However, in these previous works, the stochastic variations in micro-morphology were not investigated or quantified. In addition, the output microstructure was fixed at a certain size and was not scalable to arbitrary sizes. Furthermore, the above-mentioned methods lacked the capacity to control micro-morphology at different local regions to produce spatially varying morphologies in a single \synth{} sample. In this paper, we develop a flexible and versatile algorithm for generating realistic microstructures using GAN. The new algorithm allows control of micromorphology in different regions and can be scaled to arbitrary image dimensions seamlessly.


\begin{figure}[t]
\centering
\includegraphics[width=0.6\linewidth]{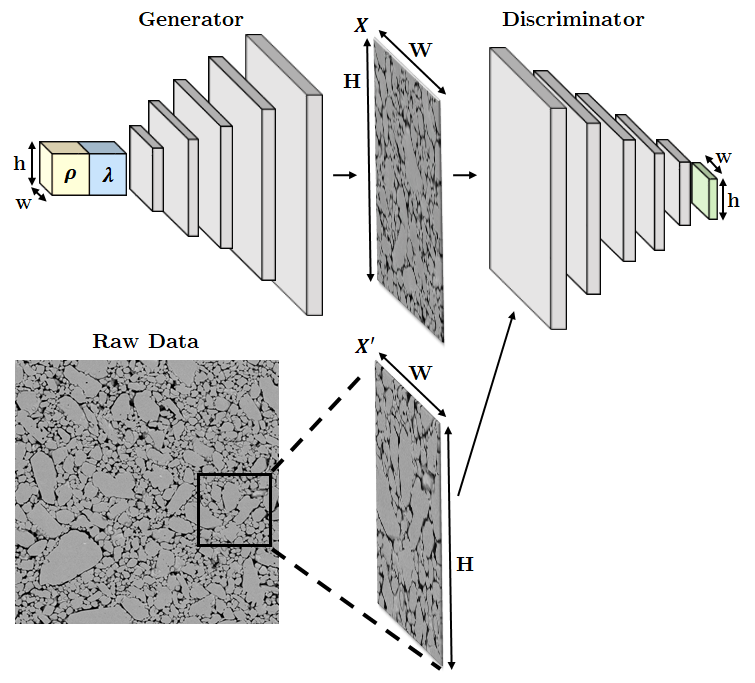}
\caption{A schematic overview of the proposed architecture.
The generator takes \textit{local stochasticity parameters} $\rho$ and the \textit{global morphology parameters} $\lambda$ as inputs, where w and h are width and height of the spatial dimension. The discriminator takes in a shuffled distribution of the generated and real microstructures.
}
\label{fig:Architecture}
\end{figure}

\subsection{A method for microstructure generation using GAN}

Here, we employ a patch-based, fully convolutional GAN architecture as illustrated in Figure~\ref{fig:Architecture}. The generator takes two input vectors $\rho \in \R^r$ and $\lambda \in \R^l$ defined at each of $h \times w$ grid locations, forming an $h \times w \times (r+l)$ input tensor. The input tensor is then up-convolved five times, each time scaling the dimension by a factor of 2, to produce an $H \times W$ microstructure image. The generator is trained together with a detector network symmetric to the generator, in which the upconvolution layers are replaced by regular convolution layers of the same kernel size and stride. During training, the detector network is presented with an arbitrarily chosen image, either \real{} or \synth{}, and is tasked to determine if the presented image is real or synthetic. The determination of whether a presented image is real or synthetic occurs at each of the $h \times w$ grid locations of the generator. The patch-wise feedback on the image quality promotes details to be more closely captured.\cite{Jetchev_2016} Furthermore, since the receptive fields of the detector network overlap by 32 pixels between adjacent patches, smooth and seamless connection between the patches is naturally enforced. Finally, it is worthwhile to note that the proposed architecture is fully convolutional so that arbitrary-sized images can be produced without stitching \cite{fokina_2019} by varying the size of the input tensor.

Two input vectors shown in Figure~\ref{fig:Architecture} ---the \textit{local stochasticity parameters} $\rho$ and the \textit{global morphology parameters} $\lambda$---play a critical role in the proposed GAN model. The role of the local stochasticity parameters $\rho$ is the same as the ``noise tensors'' in the standard GAN implementation; they serve as seeds for adding stochastic variations. The global morphology parameters $\lambda$, on the other hand, control the overall morphological characteristics of the generated image, such as grain (or void) sizes, orientations, and aspect ratios. We achieve such a control of the global morphology by setting $\lambda$ to be constant across different grid locations during training, while $\rho$ varies randomly across grid locations. Both $\rho$ and $\lambda$ are uniformly distributed in the range $[-1, 1]$ and hence parameterize the morphological variations of a material in the domain $[-1, 1]^{(l+r)}$. In the following section, we demonstrate the fidelity and versatility of the proposed GAN approach for \synth{} generation in comparison with the current state-of-the-art transfer learning (TL) approach presented Li \etal{}\cite{Li_2018}.

\begin{figure}[t]
\centering
\includegraphics[width=0.8\linewidth]{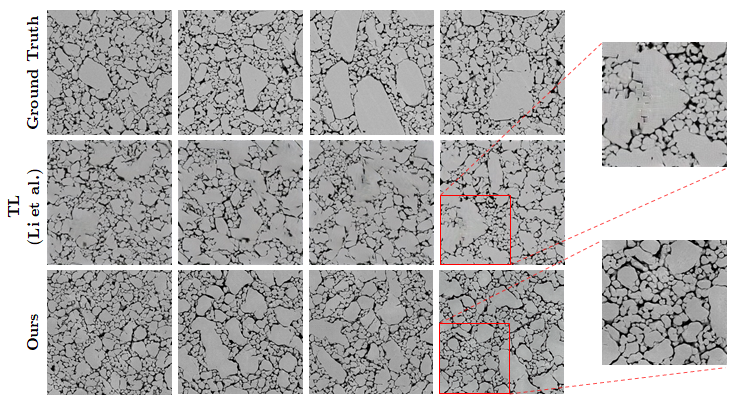}
\caption{Comparison of \real{} (top row) and two methods for generating \synth{} (middle row - TL method; bottom row - GAN). Both the benchmark TL method (Li \etal{}\cite{Li_2018}) and the proposed GAN method display realistic image quality overall. However, as seen in the image callouts to the right, small artifacts and blurry boundaries are noticeable in the benchmark method while the GAN method does not produce artifacts.}
\label{fig:Quality_comp}
\end{figure}

\section{Results}

\subsection{Qualitative and quantitative comparison of real and synthetic microstructures}

When mechanical pressing techniques are used to produce HE materials, defects such as voids, cracks, and inclusions are created.\cite{molek2017microstructural} Figure~\ref{fig:Quality_comp} (Top row labeled 'ground truth') shows examples of one such microstructure \cite{Private_2018}---sub sampled from an image obtained using SEM---illustrating the distribution of features. In this type of pressed HE, both inter-crystal and intra-crystal voids are nearly uniformly presented within the microstructure, \ie{} crystals and voids do not appear to vary significantly in size and concentration within each sampled image nor across the different samples. As shown in previous work,\cite{rai2015} when the pressed HE material is subject to shock loading in the range of 10--20 GPa, voids in the microstructure collapse, leading to localization of energy and formation of high temperature hot spots. The ignition and subsequent growth of these high temperature hot spots depends on the physicochemical properties of the crystalline material and on the morphology of the void and crystal distributions in the microstructure.\cite{springer2018, molek2020impact, springer2017effects} In the current paper, to produce \synth{} that mimic the behavior of the real (\ie{} imaged) ones, we focus on comparing the distributions of key morphological metrics of void and crystal phases and the shock response of the \real{} and \synth{}. 

\begin{figure}[t]
\centering
\includegraphics[width=\linewidth]{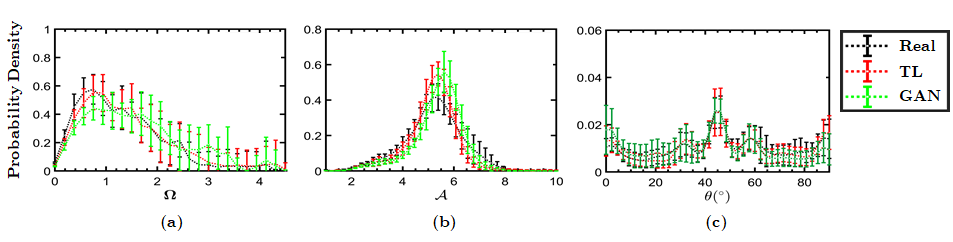}
\caption{Distributions of the morphometric parameters indicate the \synth{} are statistically similar to \real{}. The probability distribution functions (PDF) in each case (\real{}, TL-\synth{} and GAN-\synth{}) were computed for 25 sample images of size $25\mu m \times 25\mu m$. The curves in different colors indicate the mean PDFs across the 25 images, while the whiskers represent the standard deviations.}
\label{fig:Morphometry}
\end{figure}

The \real{} (\textit{top row}), \synth{} generated by the TL method of Li \etal{} \cite{Li_2018} (\textit{middle row}), and \synth{} generated by the proposed GAN method (\textit{bottom row}) are presented in Figure~\ref{fig:Quality_comp} for qualitative comparison. Overall, the GAN generated \synth{} are visually more similar to the \real{}, whereas artifacts and blurry crystal boundaries are observed in \synth{} generated by the transfer learning method. Quantitatively, we compare the statistical distributions of void shape descriptors such as void diameter $D_\text{void}$, void aspect ratio $AR$, and void orientation $\theta$, which are shown in Figure~\ref{fig:Morphometry}. From each of the categories (\ie{}, \real{}, transfer learned (TL) \synth{}, and GAN-\synth{}), 25 random samples were drawn. Morphometric analysis was performed using methods described in Roy \etal{} \cite{roy2019}. The distribution of the shape descriptors of the \real{} and \synth{} were generally in good agreement, both between the GAN-\synth{} and \real{} as well as between the TL-\synth{} and \real{}. The whiskers in the figures indicate the standard deviation among the 25 microstructures while the curves correspond to the means. In general, the generated \synth{} have voids of sizes in the same range as the \real{}; the peak of the distribution is shifted slightly in the generated microstructures. As demonstrated below, the observed differences in the size distribution have negligible effects on the computed quantity of interest, \viz{} the hot spot ignition and growth rate. The void aspect ratio distribution of the real and imaged microstructures is likewise in good agreement, albeit with a shift in the peak of the distribution. On the other hand, the void orientation $\theta$ distribution plots show good agreement; a small peak is observed at $\theta=45^\circ$ but the overall distribution of void orientations is fairly uniform, \ie{} there is no strong orientational preference of the voids. As seen from the figure, transfer learning also shows overall good agreement with the \real{} void distribution.

\begin{figure}[t]
\centering
\includegraphics[width=\linewidth]{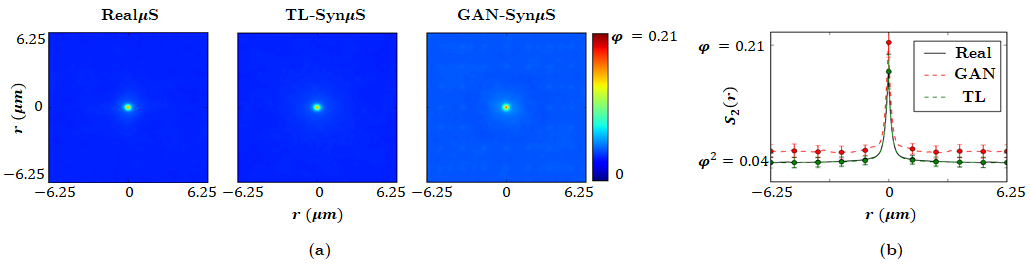}
\caption{(a) Two-dimensional two-point correlation functions for \real{} and the two different \synth{}. (b) Comparison of the cross-sectional slices of the two-dimensional two-point correlation functions in (a).}
\label{fig:two_point}
\end{figure}

In addition to the above shape descriptors, two-point correlation functions \cite{Torquato_2006} were obtained to quantify the void phase morphology. The computed two-dimensional two-point correlation functions for the \real{} and \synth{} are displayed in Figure~\ref{fig:two_point} where $\varphi(r)$ indicates the volume fraction of void phase at distance $r$. The result shows that all three microstructures have similar volume fraction of the void phase. However, unlike the transfer learning approach, the GAN generated microstructure shows a weak correlation that persists across the plot. Nevertheless, the magnitude of the correlation is small, and does not exert any discernable influence on the hot spot dynamics simulated below. Furthermore, the horizontal cross-sections of $\varphi_2(r)$ extracted from the center of the two-dimensional plots show that both the \synth{} are within the standard deviation of the \real{}. In addition, both plots stabilize to $\varphi^2$ at $r=1\mu$m, showing strong agreement of the correlation length with the \real{}. Therefore, the two-point correlation function plots reveal that the generated microstructures closely resemble the statistics of the void phase in the \real{}. Figure~\ref{fig:two_point} indicates that the TL approach produces a 2-point correlation in better agreement with the real image that does GAN. However, we remark that the performance of TL comes at the expense of solving an optimization problem for each \synth{} generation, as TL generates a \synth{} by minimizing the style difference of the generated \synth{} from a reference \real{} input. Furthermore, it should also be noted that a TL-\synth{} tends to hew close to the immediate neighborhood of the reference \real{} in the morphology space due to the style-difference minimization scheme. Whereas, the proposed GAN method can explore the entire morphology space by parameterizing the space by varying the input parameter $\lambda$, as will be demonstrated below. On the other hand, it can be argued that TL-\synth{} display slightly better agreement with the \real{} in morphometry due to the style difference minimization by which the TL method generates images: the TL approach generates \synth{} only within a narrow margin of style difference and, therefore, TL-\synth{} are biased towards the \real{} samples; whereas the proposed GAN method generates a broader range of novel microstructures. Hence, for the challenge of generating a large ensemble of \synth{} from a small set of \real{} images, the proposed GAN method offers greater practical benefits than the TL approach. Aside from the lack of artifacts and blurring in the GAN-\synth{} compared to the TL-\synth{} samples as shown in Figure~\ref{fig:Quality_comp}, the capacity to generate a broader range of novel microstructures using only a small number of \real{} samples in the case of GAN is a meaningful advantage given the typical paucity of available microstructure images for a general HE. The GAN approach therefore provides greater flexibility and versatility in the generated microstructures and is better aligned with our goal of creating a pathway to materials-by-design.

\begin{figure}[t]
\centering
\includegraphics[width=\linewidth]{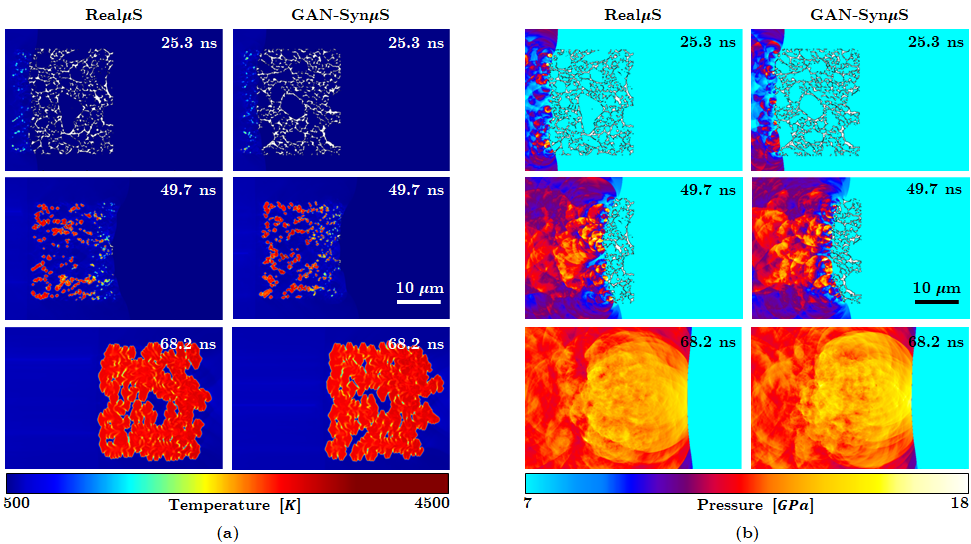}
\caption{Meso-scale shock simulations performed on \real{} and \synth{}. The \synth{} is generated to resemble the stochastic morphology of the \real{}. Note they exert similar reactive behaviors.}
\label{fig:simulation_comp}
\end{figure}

\begin{figure}[t]
\centering
\includegraphics[width=0.3\linewidth]{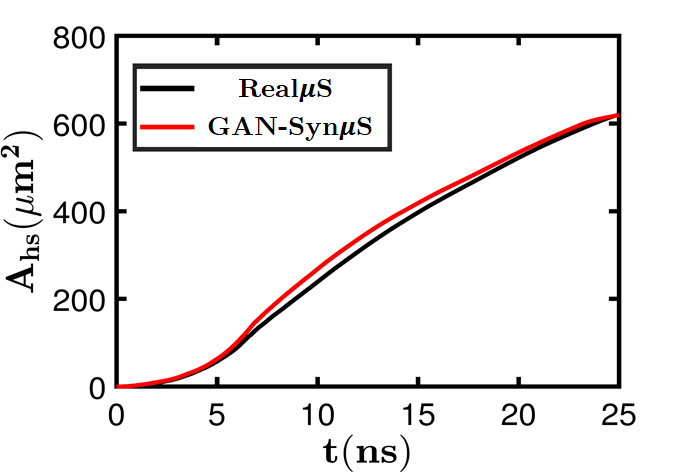}
\caption{Hot spot area evolution of \real{} and \synth{} with respect to time corresponding to the shock simulation in Figure~\ref{fig:simulation_comp}. Notice the alignment of the trend between the \real{} and the \synth{}.}
\label{fig:real_GAN_hotspot}
\end{figure}

\subsection{Simulations of reactive dynamics in real and synthetic microstructures}

The void microstructure of a pressed energetic material strongly affects the meso-scale reactive dynamics of the material under shock loading. Both experiments \cite{Price1986, molek2020impact, garcia2014shock} and simulations \cite{springer2018,levesque2015effect,rai2018three} have shown that crystal and void size distributions, void volume fractions, and void morphology affect the meso-scale sensitivity of the material. Here, reactive computations are used to compare the dynamics of the void collapse process and subsequent chemical decomposition of the solid HMX material for real and GAN generated microstructures. The methods for computation of void collapse process and reaction initiation from imaged microstructures are described in several previous publications \cite{rai2015, rai2017high, rai2018three} and only briefly outlined in the methods section of this paper.

The simulations are performed by applying shock loading from the left end of the domain at a shock pressure $P_s= 9.5$ GPa. The size of the \real{} and GAN-generated \synth{} are both $25 \mu \text{m} \times 25 \mu \text{m}$. The sample is placed in the middle of the computational domain to avoid edge effects from the domain boundaries. Extra padding regions are provided surrounding the imaged sample to account for the translation of the material during the shock passage through the material. In Figure~\ref{fig:simulation_comp}, snapshots of the evolving temperature and pressure fields are shown at several instants of time. The temperature field measures the intensity of hot spots that resulted from the process of void collapse and the pressure plot shows the blast waves that emanate from the void collapse events. As seen in the figure the temperature and pressure fields generated due to shock passage are nearly identical for the \real{} and GAN-\synth{}. Thus, the dynamics of the void collapse and hot spot evolution is well represented by the GAN-generated microstructure. A more quantitative assessment of the physically realistic response of the \synth{} is obtained from Figure~\ref{fig:real_GAN_hotspot}, which shows the time evolution of a quantity of interest in meso-scale simulations, which is the total area occupied by hot spots in the domain, calculated using the approach described in the methods section. As seen in the figure the total hot spot area $A_\text{HS}$ calculated for the \real{} and \synth{} cases is in good agreement over the course of the evolution of the hot spots. In summary, from Figure~\ref{fig:simulation_comp} the temperature and pressure fields as well as the quantity of interest (QoI) obtained from simulations using the synthetic microstructure captures the dynamics of real, imaged microstructures with good fidelity. Therefore the GAN-\synth{} can serve as suitable proxies in ensemble simulations to extract the physics of this type of pressed HMX material.


\subsection{Towards materials-by-design: Controlling the micro-morphology}

\begin{figure}[t]
\centering
\includegraphics[width=\linewidth]{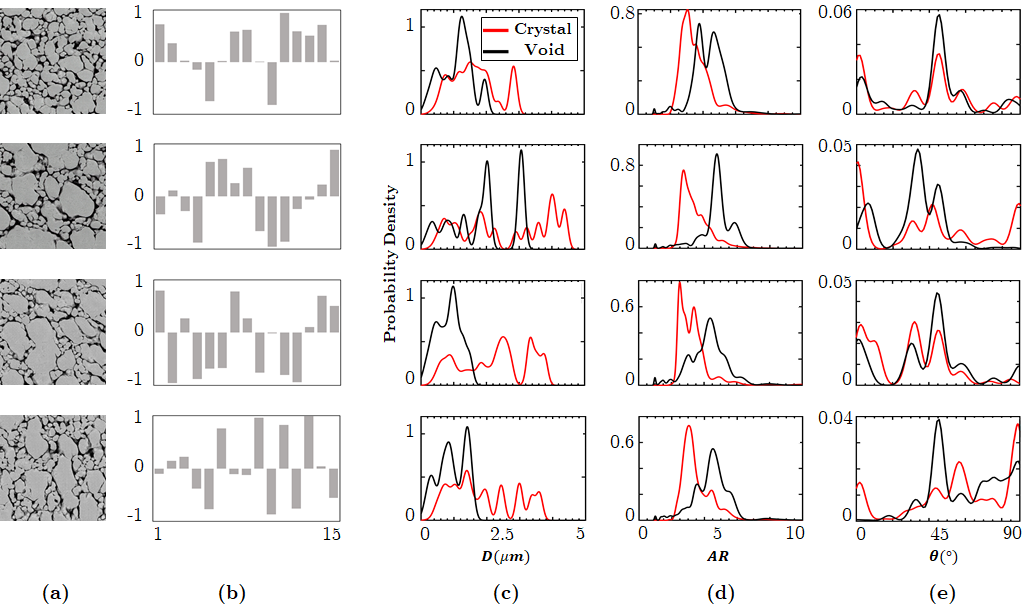}
\caption{GAN input parameter $\lambda$ controls the morphology of \synth{}. The \synth{} in column (a) were generated from $\lambda$ parameters plotted in column (b). The local stochasticity parameters $\rho$ were randomly drawn from the uniform distribution in $[-1, 1]$. Columns (c)--(e) show the morphometry of the generated images, from which the shift of distribution is noticeable as $\lambda$ varies.}
\label{fig:controllability}
\end{figure}

\paragraph{Controlling morphological features by changing $\lambda$ and $\rho$} As discussed above, the GAN model encodes global micro-morphology characteristics into a parameter vector $\lambda$. The numerical values of the elements of the vector $\lambda$ are the ``knobs,'' which can be tuned to control the morphology of \synth{}.
For example, by varying the values of $\lambda$ microstructures with different crystal morphologies are generated in Figure~\ref{fig:controllability}. The rows in the figure pertain to one set of values in the vector $\lambda$, \ie{} to one setting of the knobs that control the microstructure. The columns from left to right, in order, show the following: the generated microstructrure, the values of the elements of vector $\lambda$, the size distributions of voids and crystals in the microstructure plotted as PDFs, the distribution of void and crystal aspect ratios, and the distribution of void and crystal orientations. From visual observation as well as the morphometry plots shown alongside each image, it is clear that the parameter $\lambda$ allows for subtle control of the morphology of microstructures. For example, controlled crystal size changes are shown by comparing the first and the second rows in Figure~\ref{fig:controllability}. In the first row, the settings in the values of $\lambda$ lead to a narrow range of void and crystal sizes; by changing the values of $\lambda$ the size distribution of crystals is made wider and the void size distribution becomes strongly bimodal as seen in the second row. Meanwhile, in these two rows the void orientations and aspect ratios are largely unchanged in distribution. Controlled changes in crystal orientations are displayed in the third and fourth rows of Figure~\ref{fig:controllability}.
Due to the change of crystal orientation, the distribution of the void orientations is at the highest peak around $45^{\circ}$ in the third row, while the image in the fourth row shows void orientations predominantly at angles greater than $45^{\circ}$ with the highest value at $90^{\circ}$. In the third and fourth rows, the distribution of void and crystal sizes and aspect ratios are seen to be similar; changes in the values of $\lambda$ have primarily affected the orientations of the voids and crystals.

\begin{figure}[t]
\centering
\includegraphics[width=0.6\linewidth]{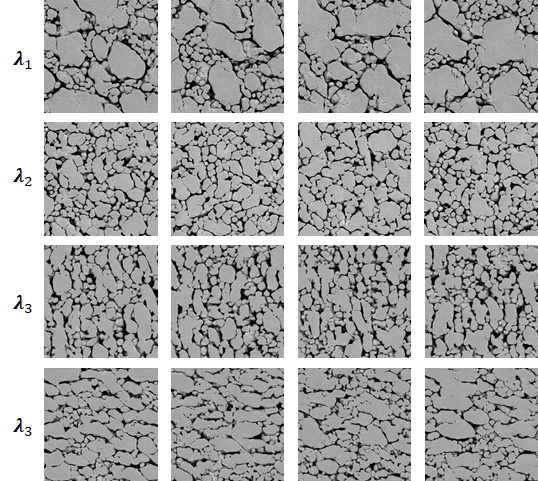}
\caption{Effects of the local stochasticity parameter $\rho$. \synth{} on the $i$-th row are generated using the same morphology parameter $\lambda_i$ but with different local stochasticity parameters. Notice the stochastic variations in morphology while the overall ``style'' is consistent.}
\label{fig:variation}
\end{figure}

Stochastic variations in the microstructure are achieved by varying the local stochasticity parameter $\rho$. Each row of Figure~\ref{fig:variation} illustrates subtle variations in micro-morphology for a specific value of the morphology parameter $\lambda$, obtained by changing the values of $\rho$. It is noticeable that the overall morphology looks similar across the columns, while the layout of crystals/voids and the subtle details are different. Note that significant variations in the void and crystal characteristics are achieved by varying the $\lambda$ (comparing across rows), while subtle variation, particularly stochasticity, is achieved by varying $\rho$ (comparing across columns).

\begin{figure}[tbp]
\centering
\includegraphics[width=0.9\linewidth]{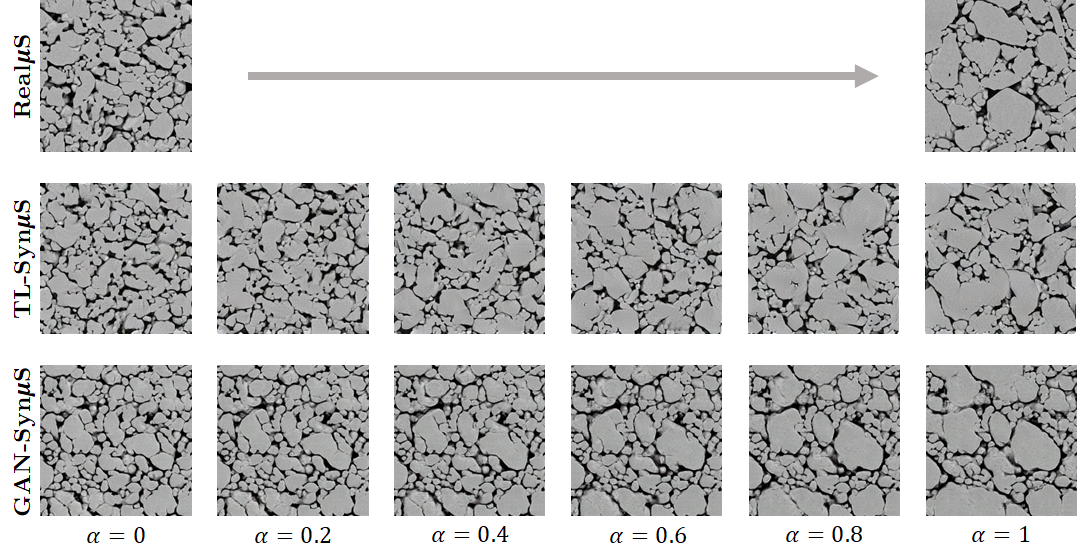}
\caption{Linear, continuous variation of micromorphology achieved via GAN. The two images on the top row display \real{} samples, and the left- and the right-most images in the middle and bottom rows display \synth{} that mimic micro-morphology of the \real{}. In between the two sides are intermediate \synth{} that are generated by linearly interpolating the Gram matrices (\textit{middle}) and the global morphology parameters $\lambda$ (\textit{bottom}). The values of $\alpha$ indicates the interpolation parameter $\lambda(\alpha) = (1-\alpha)\lambda_0 + \alpha\lambda_1$. Notice that the GAN method can model ``continuous growth'' of the crystals that is proportional to $\alpha$.}
\label{fig:interp}
\end{figure}

\paragraph{Using $\lambda$ to create "in between" and spatially varying microstructures} Another useful property of the GAN morphology parameter $\lambda$ is that changes in $\lambda$ lead to smooth and linearly proportional changes in micro-morphology, as demonstrated in Figure~\ref{fig:interp}. In the case of GAN, to span the space of morphological parameters (void/crystal size, shape, and orientation), the vector $\lambda$ was linearly interpolated between two book-ending images to produce a smoothly varying range of microstructures. In the case of TL, there was no trivial method for linearly interpolating the microstructure, but we attempted to modify the TL algorithm by making it possible to interpolate the Gram matrices (\ie{} the representation of ``styles''). Figure~\ref{fig:interp} shows that an interpolation between two different microstructures can be achieved by both TL and GAN approaches. However, the TL method was not able to generate a smooth, continuously varying interpolation, while the proposed GAN method demonstrated the opposite: \eg{} grain sizes GAN-\synth{} in Figure~\ref{fig:interp} grow almost linearly from $\alpha=0$ to $\alpha=1$. This is a clear advantage compared to the current state-of-the-art machine learning based \synth{} generation approaches, where such smooth and continuous morphology changes cannot be easily accomplished. The parameter $\lambda$ linearly maps the space of micro-morphology, so that continuously varying $\lambda$ from one value (say $\lambda_0$) to another (say $\lambda_1$) produces smoothly and linearly varying micro-morphologies.

\begin{figure}[tbp]
\centering
\includegraphics[width=\linewidth]{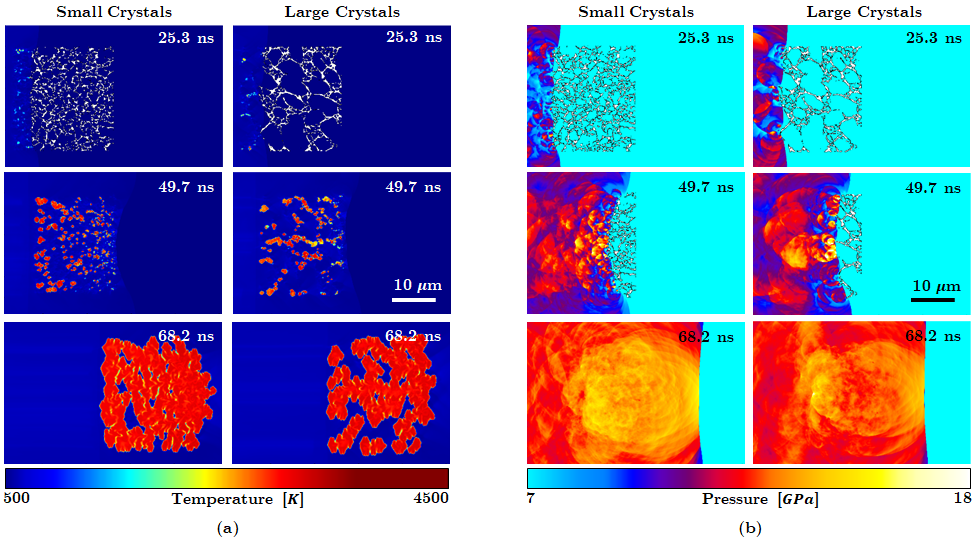}
\caption{Meso-scale shock simulations performed on controlled \synth{} containing only small crystals and another have mostly large crystals. Since morphology of the two microstructures are different, reactive behaviors are different: \synth{} with small crystals tends to increase temperature and pressure quicker than \synth{} with large crystals.}
\label{fig:simulation_control}
\end{figure}

\begin{figure}[tbp]
\centering
\includegraphics[width=0.3\linewidth]{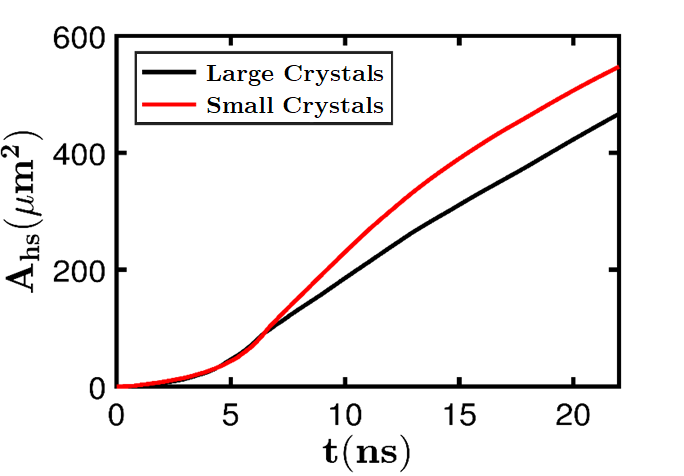}
\caption{Hot spot area evolution with respect to time. It is noticeable that the crystal size affects the growth rate of the hot spot area.}
\label{fig:hotspot_area}
\end{figure}

The ability to generate ``in-between'' morphologies using the GAN approach is of key importance for the realization of a materials-by-design framework. For example, it is well-known that the sensitivity of pressed HE samples is affected by the crystal (and concomitant void) sizes \cite{molek2020impact}. However, in previous numerical simulations of the meso-scale shock response \cite{rai2015}, it has not been possible to thoroughly quantify the behavior of a range of real, \ie{} imaged microstructures of a specific material nor the change in sensitivity due to variabilities in the microstructure; this is due to the lack of sufficiently large numbers and types of imaged samples and the lack of ability to control micro-morphology in image-based shock simulations. Using the present capability, it becomes possible to conduct shock simulations with ensembles of stochastic, morphology-controlled \synth{}. Figure~\ref{fig:simulation_control} shows results from shock simulations conducted on two samples of \synth{} with significantly different crystal/void sizes; in Figure~\ref{fig:simulation_control}~(a) the temperature field for a sample with small crystal/void sizes is shown on the left and for a sample with larger crystal/void sizes is shown on the right. In Figure~\ref{fig:simulation_control}~(a), while the difference in the mean crystal size is rather modest (around 1 $\mu$m for the left column and 2.5$\mu$m for the right column), the simulation results show significant differences in the hot spot development and therefore in the overall material sensitivity. The sample with smaller crystals/voids (left column) shows a higher density of hot spots in the control volume when compared to the sample with larger crystals/voids (right column). Figure~\ref{fig:simulation_control}~(b) shows that there is a noticeable difference also in the pressure field between the two microstructures; similar to the hot spot temperature field in Figure~\ref{fig:simulation_control}~(a), it is observed that the pressure field in the case of the small crystal size sample (left column) achieves higher overall magnitude in a larger part of the domain when compared to the pressure field for the sample with larger crystal (right column). The relatively higher sensitivity of the sample with small crystals.voids is quantitatively depicted in Figure~\ref{fig:hotspot_area} which shows the evolution of the hot spot area with time. Clearly, the small crystal microstructure has a larger hot spot growth rate compared to the large crystal microstructure.

\begin{figure}[tbp]
\centering
\includegraphics[width=\linewidth]{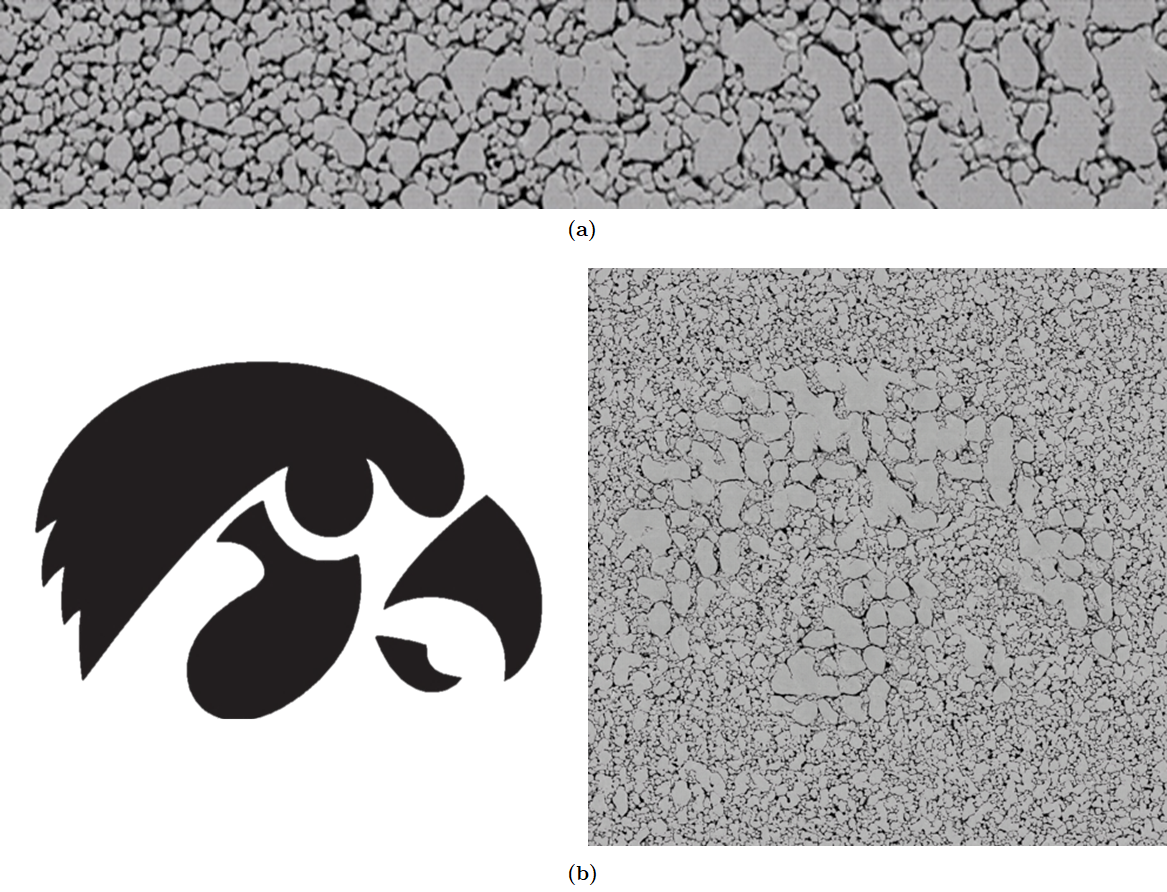}
\caption{(a) A graded microstructure generated by linearly interpolating two global morphology parameters. A global morphology parameter $\lambda_0$ corresponding to small crystal sizes and a parameter $\lambda_1$ corresponding to large crystals were obtained manually. The image was then generated by spatially grading $\lambda_0$ and $\lambda_1$ linearly: $\lambda(x) = (1-x)\lambda_0 + x\lambda_1$, where $x \in [0,1]$ is the horizontal position on the image as a fraction of the image width. (b) A layout of global morphology parameters (\textit{left}) and the corresponding GAN-\synth{}. Foreground regions were ``painted'' with $\lambda_1$ in panel (a) while the background was painted with $\lambda_0$.}
\label{fig:hawkeye}
\end{figure}

In addition to its ability to create ``in between'' microstructures by smoothly interpolating between two microstructures that define a range of possible morphologies, the proposed GAN method can exercise spatial control of the micro-morphology, providing the ability to spatially grade morphologies within a sample. This is achieved by spatially varying the morphology parameter $\lambda$. Although $\lambda$ is constrained to be constant across the spatial grid during training, at the inference time, different locations can have different values of $\lambda$. Through this, for example, one can generate a smoothly graded \synth{} by allowing $\lambda$ to vary continuously from one location to the other, as in Figure~\ref{fig:hawkeye}~(a). Here, we show that the generated \synth{} has crystal and void sizes that vary spatially, with small voids/crystals on the left end of the sample transitioning smoothly to larger voids/crystals on the right end. This type of spatial variation in the microstructure can allow for more precise control on the rate of burning of the energetic material. Furthermore, one can ``paint'' micro-morphologies by placing different ``shades'' of $\lambda$, as demonstrated in Figure~\ref{fig:hawkeye}~(b), where a "Hawkeye" pattern shown on the left is imprinted into the crystal size distribution, resulting in a microstructure on the right with a morphology that reflects this pattern. In both examples, it is noteworthy that there are no observable stitching boundaries or awkward image transitions. Using this facility, combined with the on-going micro-scale additive manufacturing (or micro 3D printing) techniques,\cite{muravyev2019, goodman2017,mcclain2019, chandru2018additive} the present controlled micro-morphology generation technique can lead to the realization of microstructure-engineered materials, which is being pursued by the authors in ongoing work.




\section{Discussion}
A novel GAN model for \synth{} generation has been developed in this work. The method is shown to produce microstructures that qualitatively and quantitatively replicate real microstructures obtained from images. The approach also provides control over the generated morphology, including the ability to spatially vary the morphological properties seamlessly. Compared to the recent deep learning based material synthesis methods,\cite{Li_2018, Cang_2018, Guo_2018, fokina_2019, Yang_2018, Mosser_2017} the advantage of the present method is the ability to scale to arbitrary size without stitching or quilting, to produce a linear and continuous control of morphology, and an ability to generate a microstructure with spatially controlled morphology.

There are several avenues for further development of the current approach. For example, the current morphology parameter is at a single scale, and cannot yet capture a large variation of crystal sizes, which occurs frequently in natural images. As described in depth in the methods section, an input grid location of the generator controls $125\times 125$ pixel area of the generated image. We found that the GAN was not able to generate crystals bigger than 125 pixels in size, which currently limits our capacity to cover the full range of morphometry. To this end, multi-scale GAN architectures recently proposed in GAN literature (\eg{}, Karras \etal{}\cite{Karras_2019}), in which the input parameters are injected at different scales of convolution are being explored. Another interesting direction of study being currently pursued is to unravel the intuitive, semantic meaning of the global morphology parameters $\lambda$. This can potentially be achieved via correlation analyses between the morphology parameters $\lambda$ and the conventional morphometric measures, such as void/crystal diameters, aspect ratios, etc. It is also remains to establish the optimal dimension of $\lambda$. Unfortunately, theoretical works on this issue are lacking in the literature, other than na\"ive search methods for finding the minimal loss by varying the dimensions. Therefore, more fundamental research should be directed on this avenue.

With regard to design and analysis of energetic materials, an immediate extension of this work would be to use the GAN method to extensively study the sensitivity of heterogeneous energetic materials to applied loads; for this purpose we will characterize the effects of micromorphology on different physical QoIs, such as hot spot ignition and growth rates \cite{rai2019void}. This can be accomplished by performing computational simulations on a large ensemble of morphology-controlled microstructures, as briefly demonstrated in Figures~\ref{fig:simulation_control} and \ref{fig:hotspot_area}. This will lead to a comprehensive quantitative characterization of the detonation behavior of HE materials as a function of micro-morphology, and will provide surrogate models for bridging meso-scale behaviors to macro-scale simulations.\cite{sen2018multi} Further development will lead to a materials-by-design framework where a desired performance or property of a HE material can be engineered through optimization of material morphology, analogous to topology optimization in mechanical component design. The ability of the proposed GAN method to spatially control micro-morphology will play a key role in the realization of such a framework.

\section{Methods}
\subsection{Data}
We used a raw image of a class V cyclotetramethylene-tetranitramine (HMX) pressed energetic material \cite{molek2017microstructural,Private_2018} obtained using SEM for the training of the GAN model. The original image was $3,000\times 3,000$ in size, from which we sampled $161\times161$ image patches by cropping the image at a random position. The physical resolution of the image is 52 nm per pixel, or 19 pixels per 1 $\mu$m. The $161\times161$ sample patches, therefore, correspond to $8.5\mu\text{m}\times 8.5\mu\text{m}$ in physical dimension.
For training, a total of 12,500 of such samples were generated and the pixel values were re-scaled from [0, 255] to [-1, 1].

\subsection{GAN}
GAN\cite{GAN_2014} is a generative model capitalizes an adversarial training of two or more competing neural networks. Typically, an image generator network $\G$ and a real/synthetic discriminator $\D$ are trained together, which poses a minimax game where $\G$ tries to maximize the misdetection rate while $\D$ tries to minimize it:
\begin{equation}
    \min_\G \max_\D V(\D, \G) =  \mathbb{E}_{x\sim p_{\text{data}}(x)} \left[ \log \D(x) \right] + \mathbb{E}_{z\sim p_z(z)} \left[ \log \left( 1-\D\left(\G(z)\right) \right) \right],
    \label{eqn:gan_loss}
\end{equation}
where $P_\text{data}$ is the distribution of real images and $z$ is the input parameter to $\G$. The convergence is reached at the equilibrium, where the discriminator $\D$ is no longer able to distinguish synthetic images generated by $\G$ from real images drawn from the data set.

In our work, the GAN loss in Equation~\ref{eqn:gan_loss} is evaluated multiple times at grid patches and averaged over grid locations, resulting the following loss function:
\begin{equation}
    \min_\G \max_\D V(\D, \G) = \frac{1}{wh}\sum_{j=1}^{h} \sum_{i=1}^{w} \mathbb{E}_{x\sim p_{\text{data}}(X)} \left[ \log \D_{j,i}(x) \right] + \frac{1}{wh}\sum_{j=1}^{h} \sum_{i=1}^{w} \mathbb{E}_{z\sim p_z(z)} \left[ \log \left( 1-\D_{j,i}\left(\G(z)\right) \right) \right],
\end{equation}
where $\D_{j,i}$ is the prediction of $\D$ at $j$-th row and $i$-th column on the $h \times w$ grid in Figure~\ref{fig:Architecture}.

The input tensor $z\in\R^{h\times w\times (r+l)}$ is composed of the local stochasticity parameter $\rho_{j,i}\in\R^r$ and the global morphology parameter vector $\lambda_{j,i}\in\R^l$ defined at each grid location $(j,i)$. During the training, we set the global morphology parameter $\lambda_{j,i}=\lambda$ for all $i, j$ for some constant randomly drawn from the uniform distribution in [-1, 1], while the local stochasticity parameter $\rho_{j,i}$ is independently drawn from the uniform random distribution in [-1, 1]. At the inference time, $\lambda_{j,i}$ can vary across the grid, as demonstrated in Figure~\ref{fig:hawkeye} and controlled by the user.

The architectures of the generator and the discriminator are symmetric. The discriminator contains a stack of five convolution layers of kernel size $5\times 5$ and stride 2. The generator has the mirrored architecture, where the convolutions are replaced by upconvolutions of the same kernel size and the stride 1/2. Such architecture results in a $125 \times 125$ receptive field, which determines the window size that one grid parameter $\lambda_{j,i}$ controls.

For the training, the adaptive moment estimation (ADAM) optimizer with the learning rate of 0.0002 and the batch size of 10 was used. We set $r=30$ and $l=15$ as the dimensions of $\rho$ and $\lambda$, respectively. During the training, $h=w=5$ was used.

\subsection{Morphometry}

Morphometry refers to the quantification of the microstructural features using shape descriptors or correlation functions. The shape descriptors calculated in this paper are used to quantify the size, shape and orientation of voids and crystals, the two primary features in the microstructure of pressed HMX. In the present work, the shape descriptors are obtained using level-set based morphometry techniques described in detail in Roy \etal{} \cite{roy2019} Briefly, the gray-scale microstructural images are segmented in the image (pixel) space using active contouring to obtain level-set fields.\cite{sethian1999level} With the zero level set value defining the crystal shapes, various approaches have been developed in Roy \etal{} \cite{roy2019} to calculate the sizes of voids and crystals, their aspect ratios and their orientations. These quantities are computed for each void/crystal in the domain. Then histograms (pdfs) are developed to obtain the distribution of void sizes, aspect ratios and orientations of all voids/crystals in the domain. The second approach to morphometry employed here is the use of correlation functions; $n$-point correlation functions provide rich information on the structures embedded in the domain. Typically, $n<3$ in the interest of computational cost. Two-point correlation functions \cite{Torquato_2006} determine the probability of finding 2 random points  with position vectors $p$ and $p+r$ in a given phase. Geometrically, the two-point correlation function $S_2(r)$ can be interpreted as the probability of having both ends of the chord vector $r$ in the same phase. The two-dimensional two-point correlation function $S_2(r)$ is calculated as the following probability function:
$S_2 (r)=P{I(p)=1,I(p+r)=1}$
where, $I(p)$ and $I(p+r)$ are the indicator function values at locations $p$ and $p+r$ respectively, $\|r\|$ is the magnitude of the length of the chord vector in the normal direction $r/\|r\|$ . The indicator function $I(p)$ is  used to identify phase at a particular point in the image space.In this work since the void phase is of interest, $I(p)$ is equal to 1 if the position vector $p$ indicates void phase and 0 otherwise. In the present work, the two-point correlations are obtained using the open source software developed by de Geus \etal{} \cite{DEGEUS2016101}

\subsection{Meso-scale simulations of reactive mechanics}
Meso-scale simulations are performed to simulate the collapse of voids in the microstructure due to the passage of a shock wave. The shock is parameterized by the pressure $P_\text{s}$. Reactive calculations are performed using methods discussed extensively in previous publications.\cite{rai2018three} Several validation exercises have been demonstrated in previous work, providing high confidence in the physical correctness of the shock computations.\cite{rai2017high} By performing the meso-scale simulations, temperature, pressure and species field data are utilized to quantify the response of the pressed material to the imposed shock, as shown in previous work.\cite{rai2015} In the present context, the QoIs used to quantify the effect of microstructure on the sensitivity of the material, are calculated by following the evolution of the temperature field  and the reaction product mass fraction in the sample. The temperature field $T(x,t)$ in the domain measures the intensity of a hot spot resulted from the process of void collapse. Higher temperature hot spots formed due the collapse of voids in the material lead to higher chemical decomposition rates. The reaction zone defines the hot spot in the domain, which is defined as the region where the temperature of the material exceeds the value of the temperature ($T_\text{bulk}$)reached after the passage of a planar shock wave. The hot spot area $A_\text{hs}$ is a significant quantity of interest for determining sensitivity and is calculated as the area of the domain where the temperature $T(x,t)>T_\text{bulk}$. The hot spot area is recorded throughout the simulation to track the evolution of the hot spots.

\section*{Acknowledgements}

This work was supported by the U.S. Air Force Office of Scientific Research (AFOSR) Multidisciplinary University Research Initiative (MURI) program (Grant No. FA9550-19-1-0318; PM: Dr. Martin Schmidt, Dynamic Materials program). We thank Drs. Chris Molek and Eric Welle from the Air Force Research Lab, Eglin AFB, for providing the images of pressed energetic materials; the images were obtained by Dr. Ryan Wixom at Sandia National Laboratories.

\section*{Author contributions statement}

H.U. and S.B. conceived the experiments. S.C., S.R., Y.N., and J.C. conducted the experiments. S.C., S.R., and Y.N. analyzed the results. All authors reviewed the manuscript.

\section*{Competing interests}

The author(s) declare no competing interests.

\bibliography{main}

\end{document}